\documentclass[letterpaper, 10pt, conference]{ieeeconf}
\IEEEoverridecommandlockouts

\usepackage{amsmath, amssymb, amsfonts}
\usepackage{amsthm}
\usepackage{mathtools}
\usepackage{bm}
\usepackage{cite}
\usepackage{graphicx}
\usepackage{xcolor}
\usepackage{booktabs}
\usepackage{algorithm}
\usepackage{algpseudocode}
\usepackage{subcaption}
\usepackage[font=small,skip=2pt]{caption}
\usepackage{titlesec}

\titlespacing*{\section}{0pt}{1.3ex plus 0.5ex minus 0.2ex}{0.8ex plus 0.2ex}
\titlespacing*{\subsection}{0pt}{1.2ex plus 0.4ex minus 0.2ex}{0.6ex plus 0.2ex}
\titlespacing*{\subsubsection}{0pt}{1.1ex plus 0.3ex minus 0.2ex}{0.4ex}

\newtheorem{theorem}{Theorem}
\newtheorem{lemma}{Lemma}
\newtheorem{definition}{Definition}
\newtheorem{remark}{Remark}
\newtheorem{corollary}{Corollary}
\newtheorem{proposition}{Proposition}
\newtheorem{problem}{Problem}
\newtheorem{assumption}{Assumption}

\newif\ifmarkrev
\markrevfalse
\ifmarkrev
  \newcommand{\rev}[1]{\textcolor{red}{#1}}
  \newenvironment{revblock}{\par\begingroup\color{red}}{\par\endgroup}
\else
  \newcommand{\rev}[1]{#1}
  \newenvironment{revblock}{}{}
\fi

\newcommand{\R}{\mathbb{R}}
\newcommand{\N}{\mathbb{N}}
\newcommand{\Rgeq}{\R_{\geq 0}}

\newcommand{\Ke}{\mathcal{K}^{e}_{\infty}}

\newcommand{\En}{\mathcal{E}^{n}_{W}}
\newcommand{\Enk}{(\mathcal{E}^{n}_{W})^{k}}

\newcommand{\Lfh}{L_{f}h}
\newcommand{\Lgh}{L_{g}h}
\newcommand{\Lfho}{L_{f}h_{1}}
\newcommand{\Lgho}{L_{g}h_{1}}
\newcommand{\psimap}{\psi_{1}(t,\mathbf{y})}
\newcommand{\KCBF}{K_{\mathrm{CBF}}}
\newcommand{\KCBFH}{K_{\mathrm{CBFH}}}
\newcommand{\KBLF}{K_{\mathrm{BLF}}}

\newcommand{\htox}{h_{0}}
\newcommand{\eps}{\varepsilon}
\newcommand{\norm}[1]{\left\lVert#1\right\rVert}
\newcommand{\innerprod}[2]{\left\langle #1,\,#2\right\rangle_{W}}

\begin{document}

\title{Enforcing a Strict Safety Margin in LLM  Text \\ Generation via Barrier Lyapunov Control}

\author{Anand Singh, Shubham Singh, Anjali Tiwari, and Katsuki Fujisawa%
\thanks{This is an extended version of a paper accepted at the Indian Control Conference (ICC) 2027.}%
\thanks{A.~Singh, S.~Singh, and K.~Fujisawa are with the Department of Computing and Mathematics,
Institute of Science Tokyo, Tokyo, Japan.
E-mail: \texttt{\{singh.a.ca69, shubham.s.ee99, fujisawa.k.2110\}@m.isct.ac.jp}.
A.~Tiwari is with the Department of Electrical Engineering,
Indian Institute of Technology Jodhpur, Jodhpur~342030, India.
E-mail: \texttt{anjali.1@iitj.ac.in}.}}

\maketitle
\begin{abstract}
Large language models are increasingly deployed without formal guarantees against toxic text generation. Existing control barrier function (CBF) based safety filters can keep the generation trajectory within a toxicity-free region, but do not enforce a positive distance from its boundary, allowing formally safe trajectories to approach the toxicity threshold arbitrarily closely and leaving little buffer against modelling errors or disturbances. We address this limitation with a Barrier Lyapunov Function (BLF) safety filter that enforces a prescribed clearance margin throughout generation while retaining the continuous-time transformer dynamics, including feedforward layers. The resulting safe input set is a half-space defined by a constraint affine in the control input, and is non-empty under certified surjectivity and classifier-nondegeneracy conditions. For margin-compatible initial conditions we prove that the clearance margin is never violated and that the prescribed clearance is restored exponentially, yielding an explicit clearance floor, and show that, as the margin vanishes, the admissible trajectories satisfy the exponential second-order CBF condition. We further derive an explicitly computable sufficient disturbance budget under which the clearance guarantee degrades gracefully. Simulations on a small transformer illustrate the resulting improvement in safety clearance.
\end{abstract}

\section{Introduction}
\label{sec:intro}
The transformer architecture~\cite{vaswani2017attention} has become
central to conversational assistants and code generation
\cite{radford2018improving}, complex reasoning and tool use
\cite{achiam2023gpt}, and open-domain question answering
\cite{guo2025deepseek}. As these models are increasingly deployed in
applications where generated content must satisfy safety requirements,
ensuring that generation avoids toxic or harmful content has become an
important challenge. Standard decoding primarily optimizes next-token
likelihood and provides no formal guarantee that the generated
trajectory remains within a desired safety region, as illustrated by
toxic generations from otherwise capable language models
\cite{gehman2020realtoxicityprompts}. This motivates safety mechanisms
that intervene during generation and provide explicit guarantees rather
than relying solely on training-time alignment or post-generation
moderation.

Existing efforts span both empirical and theoretical approaches. Studies
have investigated token-level and layer-wise representations
\cite{oh2023token}, model self-explanations~\cite{madsen2024self}, and
targeted interventions~\cite{suau2024whispering}. While effective in
practice, such approaches generally do not provide formal guarantees
throughout the generation trajectory, and recent work has shown that
safety mechanisms can be circumvented by carefully constructed inputs
and jailbreak strategies~\cite{luong2024realistic,chao2025jailbreaking}.
In parallel, continuous-time formulations have enabled a dynamical-systems
view of transformer generation~\cite{geshkovski2025mathematicalperspectivetransformers},
including analyses of clustering and consensus behavior
\cite{karagodin2024clustering,abella2025consensus}. Optimal control
\cite{cheng2025linearly} and control-barrier-function (CBF) based
filters~\cite{miyaoka2025cbfllm} have further demonstrated the potential
of control-theoretic methods for generation-time safety.

A particularly relevant formulation is~\cite{silvestre2025stopping},
which models the transformer as a continuous-time dynamical system and
intervenes through the first token to influence the final-token
representation. The resulting safety function has relative degree two
with respect to this control input, while retaining the feedforward
dynamics of the transformer. Although the resulting CBF condition
provides a formal boundary-invariance guarantee, it does not prescribe
a positive distance from the safety boundary. Thus, a trajectory may
remain formally safe while approaching the toxicity threshold
arbitrarily closely, leaving little protection against model mismatch,
classifier uncertainty, numerical errors, or small disturbances. This
reveals a gap between merely preventing boundary crossing and enforcing
a practically meaningful safety margin.

We address this gap by replacing boundary-only invariance with a
Barrier Lyapunov Function (BLF) formulation that explicitly enforces a
prescribed interior clearance from the safety boundary. We retain the
second-order, feedforward-inclusive transformer dynamics of
\cite{silvestre2025stopping}, while constructing a BLF that becomes
unbounded at the desired clearance margin. The resulting condition is
propagated through the relative-degree-two dynamics to obtain an affine
constraint on the first-token control input, enabling a minimally
restrictive quadratic-program safety filter. This formulation provides
an explicit clearance floor and a quantitative margin that can be
directly related to robustness against bounded perturbations.

The analysis is carried out on a certified operating domain where the
safety score is sufficiently smooth and the first-token input retains
nonzero control authority. These conditions are necessary for the
Lie-derivative-based BLF construction and make clear that the resulting
certificate is local to the certified model and operating region, rather
than a universal guarantee for arbitrary black-box classifiers or
deployed language models. For non-differentiable toxicity classifiers,
the formal analysis can instead be applied to a differentiable surrogate
used by the safety layer. Within this scope, the proposed framework
provides a principled step from boundary avoidance toward explicit
margin-preserving safety during transformer generation.

\textbf{\textit{Contributions.}}
The main contributions of this work are summarized as follows:
\begin{itemize}
    \item We develop a Barrier Lyapunov Function (BLF)-based safety
    filter that enforces a prescribed clearance margin $\eps$ from the
    safety boundary and restores a target clearance level $h_c>\eps$.
    The resulting admissible input set is a half-space, and the
    corresponding least-restrictive controller is the closed-form
    solution of a quadratic program.

    \item We establish non-emptiness of the safe input set under
    certified surjectivity and classifier-nondegeneracy conditions on
    the operating domain and prove that, for margin-compatible initial
    conditions, the margin is never violated. In particular, the BLF
    condition exponentially restores the desired clearance level and
    guarantees the explicit floor
    $h(t)\geq\min\{h(0),h_c\}$
    for every admissible trajectory within the certified domain. We also show that, as the clearance margin approaches zero, every admissible trajectory satisfies the exponential second-order CBF
    condition.

    \item We characterize robustness to control-channel disturbances by
    deriving a sufficient, explicitly computable disturbance budget.
    With constraint tightening, the nominal safety guarantee is
    preserved; without tightening, the certified clearance degrades
    gracefully with the disturbance magnitude. Numerical experiments
    further illustrate the clearance improvement of the proposed BLF
    filter, while the theoretical disturbance budget is left as a
    model-based certificate rather than an empirical claim.
\end{itemize}

\noindent\textbf{Notation.}
Throughout the paper, let
$r,s,k\in\N=\{1,2,\ldots\}$. Vectors in $\R^r$ are represented as
column vectors
$x=(x^1,\ldots,x^r)\in\R^r$, and a tuple of $k$ points is denoted by
$\mathbf{x}=(\mathrm{x}_1,\ldots,\mathrm{x}_k)\in(\R^r)^k$.
The space of $r\times s$ real matrices is $\R^{r\times s}$, with
$I_r$ denoting the identity matrix in $\R^{r\times r}$. For a smooth
manifold $M$ and $p\in M$, $T_pM$ denotes the tangent space at $p$,
with $X_p\in T_pM$ denoting a tangent vector. For a smooth map
$\varphi:M\to N$, its tangent map is denoted by
$T\varphi:TM\to TN$. Finally, the Lie derivative of a smooth function
$h$ along a vector field $f$ is denoted by
$L_fh = dh(f).$

\section{Problem Statement}
\label{sec:problem}

\subsection{System Description}

Following~\cite{silvestre2025stopping}, tokens evolve on the ellipsoid
$\En=\{y\in\R^{n+1}:\innerprod{y}{y}=1\}$ induced by a symmetric
positive-definite matrix $W\in\R^{(n+1)\times(n+1)}$, and a length-$k$
sequence is the tuple
$\mathbf{y}=(y_{1},\ldots,y_{k})\in\Enk$. Consider the controlled
transformer dynamics
\begin{equation}
\dot{\mathbf y}
=
f(t,\mathbf y)
+
g(t,y_1)\,u(t),
\label{eq:ct_model}
\end{equation}
where $f$ is the drift, smooth in $\mathbf{y}$ and $C^{1}$ in $t$,
collecting the self-attention and feedforward vector fields of all
tokens,
$u(t)\in T_{y_1}\En\subset\R^{n+1}$ is the control input acting on the
first token, and $g$ is the associated control vector field. Safety is
characterized by a toxicity classifier $h_0:\En\to\Rgeq$ acting on the
last token, for which we assume:

\begin{assumption}[Regularity of the toxicity classifier]
\label{ass:h0}
$\htox\in C^{2}(\En,\Rgeq)$; a less regular classifier is replaced by
a $C^{2}$ approximation.
\end{assumption}

\begin{revblock}
\textbf{Scope of the transformer model.} Model~\eqref{eq:ct_model} is the continuous-time idealization
of~\cite{silvestre2025stopping}: depth is treated as time, tokens evolve
on the ellipsoid $\En$, and the attention parameters $(P(t),U(t))$ are
given. Discrete-depth architectures, finite precision, causal masking
and sampling-based decoding of deployed LLMs are not modelled. The
control $u$ is an intervention on the first token's embedding, is
assumed unconstrained, and is assumed to be applied continuously (see
Sections~\ref{sec:blf_margin_rev} and~\ref{sec:blf_filter}). All guarantees
below are therefore statements about this model and about the score
$\htox$, not about the behaviour of a deployed LLM.

\textbf{Classifier regularity and black-box classifiers.} The $C^{2}$ requirement is used so that $h_1$ is $C^{1}$ and
$L_f h_1$ exists. Many practical classifiers (e.g.\ ReLU networks) are
not $C^{2}$, and external classifiers may be query-only. The filter
needs $d\htox$ and derivatives of $d\htox$ along $f$, so it must be run
on a differentiable surrogate $\tilde h_0\in C^{2}$ (e.g.\ a network
retrained or distilled with smooth activations such as softplus or GELU,
or a mollification of $\htox$). If
$\sup_{y\in\En}|\htox(y)-\tilde h_0(y)|\leq\eta$ and the filter enforces
margin $\tilde\eps>\eta$ for $\tilde h=\gamma-\tilde h_0$, then
$h=\gamma-\htox\geq\tilde h-\eta\geq\tilde\eps-\eta>0$: the true
classifier retains margin $\tilde\eps-\eta$, and
$\htox\leq\gamma-\tilde\eps+\eta$.  Guarantees for
non-differentiable or non-surrogable classifiers are not claimed.

\end{revblock}

Fixing a threshold $\gamma>0$, define
\begin{equation}
  h(\mathbf{y}) \;=\; -h_0(y_{k})+\gamma,
  \label{eq:h}
\end{equation}
which is $C^{2}$ under Assumption~\ref{ass:h0}, with safe set
$\mathcal{S}=\{\mathbf{y}\in\Enk:h(\mathbf{y})\ge0\}$. Since $h$
depends only on the last token while $u$ acts on the first, $h$ has
relative degree two for~\eqref{eq:ct_model}~\cite{silvestre2025stopping},
and the controller influences $h$ only through its second-order
extension
\begin{equation}
h_1(t,\mathbf{y})=(dh)_{\mathbf{y}}\bigl(f(t,\mathbf{y})\bigr)
+\alpha\bigl(h(\mathbf{y})\bigr),
\label{eq:h1_relative_degree}
\end{equation}
where $\Ke$ denotes the continuous, strictly increasing functions
$\R\to\R$ that vanish at zero and are unbounded in both directions,
and $\alpha\in\Ke\cap C^{1}(\R)$ is a design function (its
transformer-specific form is given in Section~\ref{sec:method}). Given
a clearance margin $\eps\in(0,\gamma)$, the margin-safe set is
\begin{equation}
\mathcal{S}_{\eps}=\{\mathbf{y}\in\Enk:h(\mathbf{y})\ge\eps\}
\subsetneq\mathcal{S}.
\label{eq:Seps}
\end{equation}

\subsection{Problem Formulation}

The CBF-QP framework of~\cite{silvestre2025stopping} guarantees
safety only on the bare set $\mathcal{S}$: certified trajectories may
sit arbitrarily close to $\partial\mathcal{S}$, so generated text can
be formally safe yet near-toxic, and arbitrarily small disturbances
can breach $\mathcal{S}$. Our goal is a filter enforcing the explicit
margin~\eqref{eq:Seps} while preserving CBF-QP feasibility.

\begin{problem}[Safe Text Generation via Barrier Lyapunov Control]
\label{prob:main}
Consider system~\eqref{eq:ct_model} under Assumption~\ref{ass:h0},
with margin $\eps\in(0,\gamma)$, target clearance level
$h_{c}\in(\eps,\gamma)$, decay rate $\beta>0$, a Barrier Lyapunov
Function $V$ diverging as $h\to\eps^{+}$, and a nominal first-token
controller $u^{\mathrm{nom}}:\Rgeq\times\Enk\to T_{y_1}\En$. Design a
safety-filtering law $u:\Rgeq\times\Enk\to T_{y_1}\En$ such that, for
every initial condition with $h(\mathbf{y}(0))>\eps$ in an
admissible class characterized in Section~\ref{sec:analysis}, the
closed loop satisfies, for all $t\geq0$,
\begin{align*}
\mathbf{y}(t) &\in \mathcal{S}_{\eps},\\
\dot{V}\bigl(h(\mathbf{y}(t))\bigr)
  &\leq -\beta\,\bigl(V\bigl(h(\mathbf{y}(t))\bigr)-V(h_{c})\bigr),
\end{align*}
while, at each $(t,\mathbf{y})$, $u(t,\mathbf{y})$ minimizes
$\norm{u-u^{\mathrm{nom}}(t,\mathbf{y})}^{2}$ over all inputs
satisfying the differential constraint enforcing these two conditions.
\end{problem}

The residual form of the decay condition is necessary (Section~\ref{sec:blf_margin_rev}); it certifies exponential restoration of the clearance level $h_{c}$ whenever $h<h_{c}$.

\subsection{Challenges and Research Objectives}
\label{sec:challenges}

Three challenges arise. First, the relative-degree-two safety
condition bounds only the rate of approach to $\partial\mathcal{S}$,
not the achieved distance, so trajectories may sit indefinitely close
to the boundary. Second, the constraint tightening required for a
strict margin $\eps$ must not destroy feasibility of the filtering QP.
Third, the controller acts on $h_{1}$, not on $\dot h$, so relating
the decay of $V$ to the second-order quantity $h_{1}$ requires a
dedicated argument. These motivate the research objectives:
(i)~construct a safe input set, affine in $u$, from a BLF diverging at
$h=\eps$; (ii)~prove its non-emptiness whenever the deployed
transformer satisfies the required surjectivity condition on the
certified operating domain; and (iii)~link the constraint on $h_{1}$
to the exponential restoration of the clearance level $h_{c}$, with an
explicit robustness margin.

\section{Proposed Methodology}
\label{sec:method}

We develop the BLF-based safety filter in three steps: CBF
preliminaries (Section~\ref{sec:cbf_review}), the BLF and transformer
barrier construction (Section~\ref{sec:blf_def}), and the safe input
set with its QP solution and filtering algorithm
(Section~\ref{sec:blf_filter}); guarantees follow in
Section~\ref{sec:analysis}.

\subsection{Control Barrier Functions and Second-Order CBFs}
\label{sec:cbf_review}

Consider the control-affine system
\begin{equation}
  \dot{x} = f(t,x) + g(t,x)\cdot u,
  \quad (t,x)\in\Rgeq\times\R^{n+1},
  \label{eq:sys}
\end{equation}
with $u\in\R^{m}$, $f,g$ piecewise continuous in $t$ and locally
Lipschitz in $x$, and forward-complete solutions.

\begin{definition}[Forward invariance and safety]
  A set $\mathcal{C}\subset\R^{n+1}$ is \emph{forward invariant} for
  \eqref{eq:sys} if $x_{0}\in\mathcal{C}$ implies $x(t)\in\mathcal{C}$
  for all $t\in\Rgeq$; $\mathcal{C}$ is \emph{safe} if forward
  invariant.
\end{definition}

\begin{definition}[Control Barrier Function~\cite{7782377}]
  \label{def:cbf}
  $h\in C^{1}(\R^{n+1})$ is a \emph{control barrier function} (CBF)
  for \eqref{eq:sys} on $\mathcal{S}=h^{-1}(\Rgeq)$ if there exists
  $\alpha\in\Ke$ such that, for all $(t,x)\in\Rgeq\times\mathcal{S}$,
  \begin{equation}
    \sup_{u\in\R^{m}}\bigl(\Lfh + \Lgh\cdot u\bigr)
    \;\geq\; -\alpha\bigl(h(x)\bigr).
    \label{eq:cbf_cond}
  \end{equation}
  The safe input set is
  $\KCBF(t,x)=\{u\in\R^{m}:\Lfh+\Lgh\cdot u\geq -\alpha(h)\}$.
\end{definition}

By~\cite[Theorem~2]{7782377}, $k(t,x)\in\KCBF(t,x)$ renders
$\mathcal{S}$ safe; a sufficient condition for~\eqref{eq:cbf_cond} is
$\Lgh(t,x)\neq 0$ on $\Rgeq\times\mathcal{S}$.

When $\Lgh\equiv 0$, higher-order CBFs~\cite{7524935} extend the
framework. For $h\in C^{2}(\R^{n+1})$ and
$\alpha\in\Ke\cap C^{1}(\R)$, define
\begin{equation}
  h_{1}(t,x) \;=\; (dh)_{x}\bigl(f(t,x)\bigr) + \alpha\bigl(h(x)\bigr),
  \label{eq:h1}
\end{equation}
and let $\mathcal{S}_{1}=\{x:h_{1}(0,x)\geq 0\}$.

\begin{definition}[Second-order CBF, adapted from~\cite{9516971}]
  \label{def:hocbf}
  $h\in C^{2}(\R^{n+1})$ is a \emph{second-order CBF} for
  \eqref{eq:sys} if there exist $\alpha\in\Ke\cap C^{1}(\R)$,
  $\alpha_{1}\in\Ke$, and open
  $\mathcal{D}\supset\mathcal{C}=\mathcal{S}\cap\mathcal{S}_{1}$ such
  that $\Lgh\equiv 0$ and, for all $(t,x)\in\Rgeq\times\mathcal{D}$,
  \begin{equation}
    \sup_{u\in\R^{m}}\bigl(\Lfho + \Lgho\cdot u\bigr)
    \;\geq\; -\alpha_{1}\bigl(h_{1}(t,x)\bigr).
    \label{eq:hocbf_cond}
  \end{equation}
  The safe input set is
  $\KCBFH(t,x)=\{u:\Lfho+\Lgho\cdot u\geq -\alpha_{1}(h_{1})\}$.
\end{definition}

By~\cite[Theorem~1]{9516971}, $k(t,x)\in\KCBFH(t,x)$ renders
$\mathcal{C}$ safe, with sufficient condition $\Lgho(t,x)\neq 0$ on
$\Rgeq\times\mathcal{D}$ (given $\Lgh\equiv 0$).

\subsection{Barrier Lyapunov Functions and Transformer Barrier
Construction}
\label{sec:blf_def}

Barrier Lyapunov Functions enforce constraint satisfaction with a
strict interior margin~\cite{8581068,1583507}. Unlike CBFs, which only
keep trajectories inside the safe set, BLFs additionally penalize
proximity to the boundary: their value grows unbounded as the state
approaches the clearance margin, maintaining a strictly positive
buffer.

\begin{definition}[Barrier Lyapunov Function]
  \label{def:blf}
  Let $\eps\in(0,\gamma)$ and $\mathcal{S}_{\eps}$ be as
  in~\eqref{eq:Seps}. A smooth $V:(\eps,+\infty)\to\Rgeq$ is a
  \emph{Barrier Lyapunov Function} (BLF) for $\mathcal{S}_{\eps}$ if
  (i) $V(r)\to+\infty$ as $r\to\eps^{+}$, (ii) $V$ is strictly
  decreasing on $(\eps,+\infty)$, and (iii) $V(r)\to 0$ as
  $r\to+\infty$. System~\eqref{eq:sys} is \emph{BLF-safe with residual $c\geq0$} for
  $\mathcal{S}_{\eps}$ if there exist a controller $u$ and $\beta>0$
  with
  \begin{equation}
    \dot{V}\bigl(h(x(t))\bigr)
    \;\leq\; -\beta\,\bigl(V\bigl(h(x(t))\bigr)-c\bigr),
    \quad \forall\, t\geq 0.
    \label{eq:blf_decay}
  \end{equation}
\end{definition}

Condition~(iii) makes $V$ vanish far inside the safe region -- natural
here since low-toxicity outputs incur negligible penalty -- but is not
required by the standard formulation~\cite{8581068}, which needs only
(i) and positive definiteness. The essential safety property follows
from (i) and~\eqref{eq:blf_decay} alone: by the comparison lemma,
$V(h(t))\leq\max\bigl(V(h(0)),c\bigr)<\infty$, and since
$V\to+\infty$ as $h\to\eps^{+}$, trajectories cannot escape
$\mathcal{S}_{\eps}$; we use $c=V(h_{c})$ throughout
(Section~\ref{sec:analysis}). Two
canonical BLFs are used throughout: the \emph{logarithmic BLF}
\begin{equation}
  V_{\log}(h) \;=\; \log\!\left(\frac{h}{h-\eps}\right),
  \quad h>\eps,
  \label{eq:blf_log}
\end{equation}
with $V_{\log}'(h)=-\eps/[h(h-\eps)]<0$, and the \emph{power BLF} of
order $p\geq1$,
\begin{equation}
  V_{\mathrm{pow}}(h) \;=\; \left(\frac{\eps}{h-\eps}\right)^{p},
  \quad h>\eps,
  \label{eq:blf_pow}
\end{equation}
which penalises boundary proximity more heavily. Both satisfy Definition~\ref{def:blf}.

\textbf{Transformer barrier construction.} Under
Assumption~\ref{ass:h0}, the auxiliary function $h_{1}$ is retained
unchanged from \cite[Eq.~(9)]{silvestre2025stopping},
\begin{equation}
  h_{1}(t,\mathbf{y})
  \;=\;
  -(d\htox)_{y_{k}}\!\bigl(f_{k}(t,\mathbf{y})\bigr)
  \;+\;\alpha\bigl(h(\mathbf{y})\bigr),
  \label{eq:h1_transformer}
\end{equation}
with $f_{k}=f_{\mathrm{sa},k}+f_{\mathrm{ff},k}$, together with the
controlled vector-field derivative
\begin{equation}
  \psimap
  \;\coloneqq\;
  T_{y_{k}}\pi_{W}\cdot
  \frac{\partial\tilde{f}_{\mathrm{sa},k}}{\partial y_{1}}(t,\mathbf{y})
  \;:\;
  T_{y_{1}}\En\;\to\; T_{y_{k}}\En.
  \label{eq:psi}
\end{equation}
The closed-form expression for
$\partial\tilde{f}_{\mathrm{sa},k}/\partial y_{1}$ in terms of the
key/query/value matrices $(P,U)$ of the $H$ attention heads, and the
genericity of surjectivity of $\psimap$, are established
in~\cite[Lemma~3.1, Prop.~3.1]{silvestre2025stopping} and recorded
here by reference.

\begin{lemma}[{\cite[Lemma~3.1]{silvestre2025stopping}}]
  \label{lem:dfsa}
  $\partial\tilde{f}_{\mathrm{sa},k}/\partial y_{1}(t,\mathbf{y})
  =\Xi(\mathbf{y})(P(t),U(t))$ for an explicit bilinear map $\Xi$.
\end{lemma}

\begin{proposition}[{\cite[Proposition~3.1]{silvestre2025stopping}}]
  \label{prop:surj}
  For each $(t,\mathbf{y})\in\Rgeq\times\Enk$, $\psimap$ is surjective
  for almost every
  $(P(t),U(t))\in(\R^{(n+1)\times(n+1)})^{2H}$.
\end{proposition}

The proof rests on $\det\Xi(\mathbf{y})(P,U)$ being real-analytic and
not identically zero, so its zero set has Lebesgue measure
zero~\cite{mityagin2015zerosetrealanalytic}. \rev{Note that this is a
statement for each \emph{fixed} $(t,\mathbf{y})$: the exceptional set
of parameters may depend on $(t,\mathbf{y})$.} For a fixed deployed
model, surjectivity is certified directly by checking
$\det\Xi(\mathbf{y})(P,U)\neq0$ at the deployed weights, motivating:

\begin{assumption}[Surjectivity at the deployed weights]
  \label{ass:surj}
 There is an open set $\mathcal{D}\subseteq\Enk$ such that $\psimap$
  is surjective for all $(t,\mathbf{y})\in\Rgeq\times\mathcal{D}$. By
  Proposition~\ref{prop:surj}, this holds \rev{generically, i.e.\ it is
  not excluded by a measure-zero set of parameters at each
  $(t,\mathbf{y})$; for a given deployed model it must be certified on
  $\mathcal{D}$ (see Section~\ref{sec:blf_margin_rev})}.
\end{assumption}

\subsection{BLF-Based Safety Filter}
\label{sec:blf_filter}

\textbf{Required rate.} Since $V'<0$, the decay condition of
Problem~\ref{prob:main} holds whenever $\dot h\geq w(h)$ for a
required rate $w$ determined by $V$, $\beta$, and $h_{c}$; its
explicit form~\eqref{eq:wrate}, and the transmission of this
first-order requirement to the input through the second-order
quantity $h_{1}$ (as $\Lgh\equiv0$), are derived in
Section~\ref{sec:analysis}.

\textbf{Safe input set and quadratic program.} Given a bound
$b:\Rgeq\times\Enk\to\R$, the \emph{BLF-based safe input set} is
\begin{equation}
  \KBLF^{b}(t,\mathbf{y})
  \coloneqq
  \bigl\{u\in T_{y_{1}}\En:\;
  \Lfho + \Lgho\cdot u \;\geq\; b(t,\mathbf{y})
  \bigr\},
  \label{eq:KBLF}
\end{equation}
which is a closed half-space whenever $\Lgho\neq0$. With unconstrained
$u$, feasibility is guaranteed in that case; if $\Lgho=0$, feasibility
reduces to the scalar condition $\Lfho\geq b$, but there is no
instantaneous control authority over $h_1$. The bound $b$ used in the guarantees is derived in
Section~\ref{sec:analysis}. For any $b$, the least-restrictive safe
controller solves
\begin{equation}
  \min_{u}\;\norm{u-u^{\mathrm{nom}}}^{2}
  \quad\text{s.t.}\quad
  \Lfho + \Lgho\cdot u \;\geq\; b(t,\mathbf{y}),
  \label{eq:qp}
\end{equation}
with closed-form solution
\begin{equation}
  u^{*} = u^{\mathrm{nom}}
  + \max\!\Bigl(0,\,
    \tfrac{b-\Lfho-\Lgho\cdot u^{\mathrm{nom}}}
         {\norm{\Lgho}^{2}}\Bigr)
    \Lgho^{\top}.
  \label{eq:qp_soln}
\end{equation}
Evaluating $V(h)$, $w(h)$, and $b$ adds only $O(1)$ operations
per step; the dominant cost, evaluating $\Lfho$ and $\Lgho$, is
unchanged from the CBF case of~\cite{silvestre2025stopping}.

\begin{revblock}
\textbf{Computational cost and scaling.} Per step, the filter needs $h$, $h_1$, $L_f h_1$ and $L_g h_1$. With
automatic differentiation these amount to one gradient of the classifier
($d\htox$), Jacobian--vector products of $f_k$ and of $d\htox$ along
$f$, and the closed-form $\psi_1$ of Lemma~\ref{lem:dfsa}; the cost is
therefore a small constant multiple of one evaluation of the drift
(which scales with the number of tokens, the embedding dimension and the
number of heads) plus $O(1)$ for $V$, $w$, $b$ and
the closed-form QP~\eqref{eq:qp_soln}. The present discussion is a
complexity estimate for the filtering operation, not an empirical
real-time claim for pretrained GPT-2-scale models. A pretrained model
such as GPT-2 has discrete depth, layer normalization and causal masking;
casting it in the form \eqref{eq:ct_model} and running the filter on it
is beyond the scope of this paper and left to future work.

\end{revblock}

\textbf{Filtering algorithm.} Surjectivity is certified offline on the
operating domain $\mathcal{D}$ at the deployed weights via
$\det\Xi(\mathbf{y})(P,U)\neq0$ for all $\mathbf{y}\in\mathcal{D}$
(Assumption~\ref{ass:surj}). At each step, $h$, $h_{1}$, and $b$ are
evaluated and~\eqref{eq:qp} is solved via~\eqref{eq:qp_soln}. With
unconstrained $u$, feasibility is guaranteed whenever $\Lgho\neq0$;
if $\Lgho=0$, feasibility requires $\Lfho\geq b$, but there is no
instantaneous control authority over $h_1$. In that degenerate regime,
the filter passes $u^{\mathrm{nom}}$ through only when the scalar
condition holds; otherwise the certificate is unavailable. Algorithm~\ref{alg:blf} summarises the
procedure.

\begin{algorithm}[t]
\caption{BLF-Filtered Safe Token Generation}
\label{alg:blf}
\begin{algorithmic}[1]
\Require Prompt $\mathbf{y}^{0}$, threshold $\gamma$, margin $\eps$,
clearance level $h_{c}$, decay rate $\beta$, BLF $V$, functions
$\alpha\in\Ke\cap C^{1}(\R)$, $\mu\in\Ke$
\Ensure Safe trajectory $\mathbf{y}(t)$, $t\in[0,T]$
\State Certify $\det\Xi(\mathbf{y})(P(t),U(t))\neq 0$ for all
$(t,\mathbf{y})\in\Rgeq\times\mathcal{D}$ at the deployed parameter trajectory
(Assumption~\ref{ass:surj})
\State Initialize $\mathbf{y}(0)\in\mathcal{D}$ with
$h(\mathbf{y}(0))>\eps$ and
$h_{1}(0,\mathbf{y}(0))\geq\rho(h(\mathbf{y}(0)))$
\For{$t=0,\Delta t,2\Delta t,\dots,T$}
\State Evaluate $h(\mathbf{y}(t))$, $h_{1}(t,\mathbf{y}(t))$, and the
bound $b(t,\mathbf{y}(t))$ of Section~\ref{sec:analysis}
\State Compute nominal control $u^{\mathrm{nom}}(t,\mathbf{y}(t))$
\If{$\Lgho(t,\mathbf{y}(t))\neq 0$}
\State Solve QP~\eqref{eq:qp} for $u^{*}$ via~\eqref{eq:qp_soln}
\ElsIf{$\Lfho(t,\mathbf{y}(t))\geq b(t,\mathbf{y}(t))$}
\State $u^{*}\gets u^{\mathrm{nom}}$ \Comment{no instantaneous control authority}
\Else
\State certificate unavailable / infeasible
\EndIf
\State Apply $u^{*}$ to~\eqref{eq:ct_model}; advance
$\mathbf{y}(t+\Delta t)$
\EndFor
\end{algorithmic}
\end{algorithm}

\begin{revblock}
\textbf{Continuous-time guarantees vs.\ discrete-time implementation.} Algorithm~\ref{alg:blf} evaluates $u^{*}$ at the sampling instants and
holds it over $[t,t+\Delta t)$, while the guarantees of
Section~\ref{sec:analysis} are for the continuous-time closed loop in
which $u^{*}$ is updated at every instant. Between samples, the
constraint $\Lfho+\Lgho\cdot u\geq b$ is therefore satisfied only up to
an inter-sample error that, on any compact subset of
$\mathcal{S}_{\eps}\cap\mathcal{D}$ and for smooth data, is $O(\Delta t)$;
in the analysis it enters exactly like the disturbance of
Proposition~\ref{prop:robust} and can be absorbed by tightening $b$ by
 a term proportional to $\Delta t$. We do not derive the constants here:
a rigorous sampled-data certificate, and the corresponding selection of
$\Delta t$, are left to future work.
Accordingly, the formal guarantees in Section~\ref{sec:analysis}
apply to the continuous-time closed loop; the discrete implementation
is an approximation whose safety loss due to inter-sample error is not
formally certified here. In particular, the strict-margin violations observed in
Section~\ref{sec:simulation} occur in the sampled implementation, to
which Theorem~\ref{thm:blf_main} does not apply.

\end{revblock}

\section{Safety and Feasibility Analysis}
\label{sec:analysis}

This section establishes: (i) the bound $b$ completing the safe input
set $\KBLF=\KBLF^{b}$, with non-emptiness and relaxed BLF decay, hence
margin safety for margin-compatible initial conditions with an
explicit clearance floor; (ii) the relation to exponential second-order CBF
safety as $\eps\to0$; (iii) robustness under control-channel
disturbances with an explicit budget; and (iv) the overall guarantee
under the certified surjectivity and nondegeneracy conditions on the
operating domain.

\subsection{Required Rate, Safe Input Set, and BLF Safety}
\label{sec:blf_margin_rev}

Since $h\leq\gamma$ along every trajectory (as $\htox\geq0$), the
strict decay $\dot V\leq-\beta V$ is not sustainable: it would force
$V(h(t))<V(\gamma)$, i.e.\ $h(t)>\gamma$, in finite time. We therefore
fix a \emph{target clearance level} $h_{c}\in(\eps,\gamma)$ and enforce
the relaxed decay condition
\begin{equation}
  \dot V\bigl(h(t)\bigr)
  \;\leq\;
  -\beta\bigl(V(h(t))-V(h_{c})\bigr),
  \qquad \forall\,t\geq0,
  \label{eq:blf_decay_rel}
\end{equation}
which restores clearance exponentially whenever $h<h_{c}$ and imposes
no decay above $h_{c}$. Since $V'<0$, \eqref{eq:blf_decay_rel} holds
whenever $\dot h\geq w(h)$, with the \emph{required rate}
\begin{equation}
  w(h)
  \;\coloneqq\;
  \frac{\beta\bigl(V(h)-V(h_{c})\bigr)}{\lvert V'(h)\rvert},
  \qquad h>\eps,
  \label{eq:wrate}
\end{equation}
a $C^{1}$ function, positive on $(\eps,h_{c})$, zero at $h_{c}$, and
negative on $(h_{c},+\infty)$, so that above $h_{c}$ the certificate
permits $h$ to decrease at rate up to $\lvert w(h)\rvert$; the QP
constraint, which acts on $h_{1}$, may nevertheless be active at any
clearance level. Since $h$ has relative degree two, the controller acts on
$\dot h$ only through $h_{1}=\dot h+\alpha(h)$; define the target
margin function
\begin{equation}
  \rho(h) \;\coloneqq\; w(h) + \alpha(h),
  \qquad h>\eps,
  \label{eq:rho}
\end{equation}
so that $h_{1}\geq\rho(h)$ implies
$\dot h=h_{1}-\alpha(h)\geq w(h)$, bridging the second-order dynamics
and the first-order condition~\eqref{eq:blf_decay_rel}. Fixing an
auxiliary $\mu\in\Ke$ (a design choice, as $\alpha_{1}$ is for
$\KCBFH$), the safe input set of~\eqref{eq:KBLF} is completed by the
bound
\begin{equation}
  b(t,\mathbf{y})
  \;\coloneqq\;
  \rho'(h)\bigl(h_{1}-\alpha(h)\bigr)
  -\mu\bigl(h_{1}-\rho(h)\bigr),
  \label{eq:bound}
\end{equation}
evaluated at $h=h(\mathbf{y})$, $h_{1}=h_{1}(t,\mathbf{y})$, and we
set $\KBLF\coloneqq\KBLF^{b}$. The right-hand side of the constraint
is $u$-independent, so~\eqref{eq:qp}--\eqref{eq:qp_soln} apply
verbatim. Beyond $h(\mathbf{y}(0))>\eps$, initial conditions must be
\emph{margin-compatible},
\begin{equation}
  h_{1}(0,\mathbf{y}(0)) \;\geq\; \rho\bigl(h(\mathbf{y}(0))\bigr).
  \label{eq:margin_compat}
\end{equation}

Non-emptiness of $\KBLF$ requires control authority over $h_{1}$,
i.e.\ $\Lgho\neq0$; since
$\Lgho=-(d\htox)_{y_{k}}\circ\psi_{1}$~\cite{silvestre2025stopping},
this holds under Assumption~\ref{ass:surj} together with:

\begin{assumption}[Nondegeneracy of the classifier]
  \label{ass:nondeg}
   $(d\htox)_{y_{k}}\neq 0$ as a linear functional on $T_{y_{k}}\En$
  for every $\mathbf{y}\in\mathcal{D}$, where
  $\mathcal{D}$ is the open set of Assumption~\ref{ass:surj}.
\end{assumption}

\begin{remark}
  \label{rem:critical}
  As $\htox$ attains its extrema on the compact $\En$,
 Assumption~\ref{ass:nondeg} requires excluding a neighbourhood of
  the critical points of $\htox$ from $\mathcal{D}$, so
  $\mathcal{D}\not\supset\mathcal{S}_{\eps}$ in general and the
  guarantees below hold while the trajectory evolves in $\mathcal{D}$.
  At minimizers of $\htox$ with $\htox=0$ the clearance is maximal
  ($h=\gamma>h_{c}$) and $w(h)<0$, but the constraint may still be
  active there and the correction in~\eqref{eq:qp_soln} is unbounded
  as $\Lgho\to0$; for $n\geq2$, saddle points of $\htox$ may moreover
  lie inside $\mathcal{S}_{\eps}$. The guarantee is therefore local to the non-degenerate
  operating region $\mathcal{D}$; trajectories that leave $\mathcal{D}$
  are outside the scope of the certificate.
\end{remark}

\begin{theorem}[Continuous-time BLF safety condition]
  \label{thm:blf_main}
 Let $h$ be as in~\eqref{eq:h} and $h_{1}$ as
  in~\eqref{eq:h1_transformer}, and suppose
  Assumptions~\ref{ass:h0}--\ref{ass:nondeg} hold. Let
  $\mathcal{D}_{\eps}:=\mathcal{D}\cap\{\mathbf{y}:h(\mathbf{y})>\eps\}$.
  Then $\KBLF(t,\mathbf{y})$ is non-empty for every
  $(t,\mathbf{y})\in\Rgeq\times\mathcal{D}_{\eps}$. Moreover, if
  $\mathbf{y}(0)\in\mathcal{D}_{\eps}$ and~\eqref{eq:margin_compat}
  holds, any continuous controller
  $k(t,\mathbf{y})\in\KBLF(t,\mathbf{y})$
  satisfies~\eqref{eq:blf_decay_rel} and keeps the trajectory in
  $\mathcal{S}_{\eps}$ for as long as it evolves in $\mathcal{D}$; in
  particular, over that interval,
  \begin{equation}
    h\bigl(\mathbf{y}(t)\bigr)
    \;\geq\;
    \min\bigl(h(\mathbf{y}(0)),\,h_{c}\bigr)
    \;>\;\eps.
    \label{eq:floor}
  \end{equation}
\end{theorem}

\begin{proof}
  Let $\mathbf{y}(\cdot)$ be any closed-loop solution
  of~\eqref{eq:ct_model} with $u(t)=k(t,\mathbf{y}(t))$ (the QP
  solution~\eqref{eq:qp_soln} is continuous on
  $\Rgeq\times\mathcal{D}_{\eps}$; the argument below does not use
  uniqueness), and let
  $[0,T^{*})$ be the maximal interval on which $\mathbf{y}(t)\in\mathcal{D}$
  and $h(\mathbf{y}(t))>\eps$; $T^{*}>0$ since $\mathbf{y}(0)\in\mathcal{D}$
  and $h(\mathbf{y}(0))>\eps$, and $\mathcal{D}$ and $\{h>\eps\}$ are open.

  \textbf{Step 1 (non-emptiness).} The defining constraint
  The defining constraint is affine in $u$ with coefficient
  $\Lgho=-(d\htox)_{y_{k}}\circ\psimap$, which is nonzero on
  $\Rgeq\times\mathcal{D}_{\eps}$ by Assumptions~\ref{ass:surj}--\ref{ass:nondeg}.
  For any $u_{0}$, the input $u=u_{0}+\lambda\,\Lgho^{\top}$ with
  $\lambda\geq\bigl(b-\Lfho-\Lgho\cdot u_{0}\bigr)/\norm{\Lgho}^{2}$
  satisfies it, so $\KBLF(t,\mathbf{y})\neq\varnothing$ is a closed
  half-space for every $(t,\mathbf{y})\in\Rgeq\times\mathcal{D}_{\eps}$.

  \textbf{Step 2 (dynamics of $\sigma$).} Since $\Lgh\equiv0$, along
  the solution $\dot h=h_{1}-\alpha(h)$. On $[0,T^{*})$ the function
  $\sigma(t)\coloneqq h_{1}(t,\mathbf{y}(t))-\rho\bigl(h(\mathbf{y}(t))\bigr)$
  is $C^{1}$, because $h_{1}$ is $C^{1}$ ($\htox\in C^{2}$) and
  $\rho\in C^{1}(\eps,\infty)$ ($V$ is smooth). Differentiating, and
  using $\dot h_{1}=\Lfho+\Lgho\cdot u$ (with $\Lfho$ including the
  partial derivative in $t$),
  \[
\begin{aligned}
\dot\sigma&=(\Lfho+\Lgho\cdot u)-\rho'(h)\bigl(h_{1}-\alpha(h)\bigr)\\
&\geq b-\rho'(h)\bigl(h_{1}-\alpha(h)\bigr)=-\mu(\sigma),
\end{aligned}
\]
  where the inequality is the defining constraint of $\KBLF$ and the
  last equality is~\eqref{eq:bound}.

  \textbf{Step 3 ($\sigma\geq0$).} By~\eqref{eq:margin_compat},
  $\sigma(0)\geq0$. Suppose $\sigma(t_{1})<0$ for some
  $t_{1}\in(0,T^{*})$, and let
  $t_{0}\coloneqq\sup\{t<t_{1}:\sigma(t)\geq0\}$. Then
  $\sigma(t_{0})=0$ and $\sigma<0$ on $(t_{0},t_{1}]$. Since
  $\mu\in\Ke$ is strictly increasing with $\mu(0)=0$, $\mu(\sigma)<0$
  there, so $\dot\sigma\geq-\mu(\sigma)>0$ on $(t_{0},t_{1}]$ and
  $\sigma(t_{1})>\sigma(t_{0})=0$, a contradiction. Hence
  $\sigma\geq0$, i.e.\ $h_{1}\geq\rho(h)$, on $[0,T^{*})$.

  \textbf{Step 4 (decay of $V$).} From Step~3,
  $\dot h=h_{1}-\alpha(h)\geq\rho(h)-\alpha(h)=w(h)$. As $V'<0$,
  multiplying by $V'(h)$ reverses the inequality:
  \[
    \dot V(h)=V'(h)\,\dot h\leq V'(h)\,w(h)
    =-\beta\bigl(V(h)-V(h_{c})\bigr),
  \]
  by~\eqref{eq:wrate}, which is~\eqref{eq:blf_decay_rel}. With
  $z(t)\coloneqq V(h(t))-V(h_{c})$ we have $\dot z\leq-\beta z$, so
  $\tfrac{d}{dt}\bigl(e^{\beta t}z\bigr)\leq0$ and
  $z(t)\leq z(0)e^{-\beta t}$ on $[0,T^{*})$.

  \textbf{Step 5 (floor and conclusion).} Step~4 gives
  $V(h(t))\leq V(h_{c})+\bigl(V(h(0))-V(h_{c})\bigr)e^{-\beta t}
  \leq\bar V\coloneqq\max\{V(h(0)),V(h_{c})\}<\infty$. Because $V$ is
  continuous and strictly decreasing, $V(h)\leq\bar V$ is equivalent to
  $h\geq V^{-1}(\bar V)=\min\{h(0),h_{c}\}$, which is~\eqref{eq:floor}
  and is strictly greater than $\eps$. Thus on $[0,T^{*})$ the state
  remains in the closed set $\{h\geq\min\{h(0),h_{c}\}\}$, bounded away
  from $\{h=\eps\}$, so $T^{*}$ cannot end because $h$ reaches $\eps$.
  Hence $T^{*}$ is the first exit time from $\mathcal{D}$ (or $\infty$),
  and for $t\in[0,T^{*})$ we have $\mathbf{y}(t)\in\mathcal{S}_{\eps}$
  and~\eqref{eq:blf_decay_rel}, as claimed.
\end{proof}

\begin{revblock}
\textbf{Admissible initial conditions.} Since $\Lgh\equiv0$, $\dot h=h_1-\alpha(h)$, so
\eqref{eq:margin_compat} is equivalent to $\dot h(0)\geq w(h(0))$:
it restricts the initial \emph{rate} of change of the margin, not just
its value. Because $w\leq0$ for $h\geq h_c$, every state with
$h(0)\geq h_c$ that is not decreasing faster than $|w(h(0))|$ (in
particular every non-decreasing one) is admissible; closer to the
margin, $w>0$ and the state must already be improving at rate $w$. Both
$h(0)$ and $h_1(0)$ are computable from the prompt state, so
admissibility can be checked before generation starts. A prompt that
violates~\eqref{eq:margin_compat}, or has $h(0)\leq\eps$, lies outside
the guarantee; handling such prompts  is not addressed here. The
simulations of Section~\ref{sec:simulation} use margin-compatible
initial conditions only.

\textbf{Control authority and its limits.} Authority over $h_1$ comes only from $\Lgho=-(d\htox)_{y_k}\circ\psi_1$,
which is nonzero on $\mathcal{D}$ by
Assumptions~\ref{ass:surj}--\ref{ass:nondeg}; the guarantees hold only
while the trajectory remains in $\mathcal{D}$, and the pass-through
branch of Algorithm~\ref{alg:blf} offers no guarantee. The input is
assumed unconstrained: the filter's correction satisfies
$\|u^{*}-u^{\mathrm{nom}}\|\leq
\max\{0,\,b-\Lfho-\Lgho\cdot u^{\mathrm{nom}}\}/\|\Lgho\|$, which
grows without bound as $\|\Lgho\|\to0$, i.e.\ near the degenerate set
(near the critical points of $\htox$ or where $\psi_1$ loses rank). With
bounded inputs $\|u\|\leq\bar u$ the constraint may become infeasible
and the guarantees are not claimed.

\textbf{What ``almost every parameter'' does and does not mean.} Proposition~\ref{prop:surj} states, for each fixed $(t,\mathbf{y})$,
that parameters $(P,U)$ for which $\psi_1$ is not surjective form a
Lebesgue-null set. It does \emph{not} assert a single null set
independent of $(t,\mathbf{y})$; hence the guarantees are formally
conditioned on Assumption~\ref{ass:surj} holding on the open set
$\mathcal{D}$ visited by the trajectory, which for a deployed model is
certified directly (Algorithm~\ref{alg:blf}, line~1). This genericity
result should not be interpreted as an empirical claim that a pretrained
LLM necessarily satisfies Assumption~\ref{ass:surj}; for a particular
trained model, the condition must be checked on the certified operating
domain $\mathcal{D}$ before the safety guarantee is invoked. The result
also gives no non-emptiness on all of $\mathcal{S}_\eps$
(Remark~\ref{rem:critical}) and concerns attention parameters only:
Assumption~\ref{ass:nondeg} is a property of the classifier, not a
generic property of the parameters. Thus, the certificate is local to
the non-degenerate operating region rather than a guarantee over the
entire state manifold.
\end{revblock}

\subsection{Relation to the CBF Case}

\label{sec:reduction}

\begin{corollary}[Exponential CBF safety as a limiting case]
  \label{cor:limit}
  For the power BLF~\eqref{eq:blf_pow} of order $p\geq1$ with $h_{c}$
  fixed, the required rate satisfies, for every fixed $h>0$,
  \[
    w_{\eps}(h)
    \;=\;
    \frac{\beta}{p}\Bigl[(h-\eps)
    -\frac{(h-\eps)^{p+1}}{(h_{c}-\eps)^{p}}\Bigr]
    \;\xrightarrow[\eps\to0^{+}]{}\;
    \frac{\beta}{p}\,h\Bigl[1-\bigl(\tfrac{h}{h_{c}}\bigr)^{p}\Bigr].
  \]
  Since $h\leq\gamma$, the limiting enforced inequality
  $\dot h\geq w_{0}(h)$ implies $\dot h\geq-\kappa^{*}h$ with
  $\kappa^{*}=\frac{\beta}{p}\bigl[(\gamma/h_{c})^{p}-1\bigr]$, i.e.\
  $h_{1}\geq0$ in~\eqref{eq:h1} with $\alpha(h)=\kappa^{*}h$. Thus,
  as $\eps\to0^{+}$, every BLF-admissible trajectory remains in the
  invariant set $\mathcal{C}$ of the exponential second-order CBF
  of~\cite{7524935}; the converse does not hold.
\end{corollary}

\begin{proof}
  Substituting $V_{\mathrm{pow}}$ and
  $V_{\mathrm{pow}}'=-p\,\eps^{p}(h-\eps)^{-p-1}$
  into~\eqref{eq:wrate} gives $w_{\eps}$; the limit is immediate. On
  $(0,\gamma]$,
  $w_{0}(h)/h=\frac{\beta}{p}[1-(h/h_{c})^{p}]
  \geq-\frac{\beta}{p}[(\gamma/h_{c})^{p}-1]=-\kappa^{*}$.
\end{proof}

The logarithmic BLF is recovered as the $p=1$ case in the limit:
$w_{\eps}\to\beta h(1-h/h_{c})$ by
$\log\bigl(h/(h-\eps)\bigr)\sim\eps/h$. Recovering an arbitrary
class-$\mathcal{K}$ comparison rate would require a BLF family beyond
the log/power pair, and is left open.

\subsection{Robustness to Disturbances}
\label{sec:robust_formal}

Suppose disturbances enter through the control channel,
\begin{equation}
\dot{\mathbf y} = f(t,\mathbf y) + g(t,y_1)\bigl(u(t)+d(t)\bigr),
\qquad \norm{d(t)}\le\delta,
  \label{eq:perturbed}
\end{equation}
which models perturbations entering through the first-token input
channel, such as errors in the applied intervention.

\begin{assumption}
  \label{ass:disturbance_channel}
  The disturbance acts only through the control channel $g(t,y_{1})$,
  as in~\eqref{eq:perturbed}.
\end{assumption}

By Theorem~\ref{thm:blf_main}, trajectories remain in
$\mathcal{S}_{\eps}$ for as long as they remain in $\mathcal{D}$.
Assume that the control-channel gain is uniformly bounded on the
certified operating region:
\[
L_{h_1}\coloneqq
\sup_{t\geq0,\;\mathbf{y}\in\mathcal{S}_{\eps}\cap\mathcal{D}}
\|\Lgho(t,\mathbf{y})\|<\infty.
\]
This is the norm of the control-channel gain $\Lgho$, not
$\sup\|dh_1\|$: the disturbance enters $\dot h_1$ only through $g$, so
$|\Lgho\cdot d|\leq L_{h_1}\delta$ (and $L_{h_1}\leq\sup\|dh_1\|$
whenever $\|g\|\leq1$).

\begin{proposition}[Robust margin safety via tightening]
  \label{prop:robust}
   Under the hypotheses of Theorem~\ref{thm:blf_main} and
  Assumption~\ref{ass:disturbance_channel}, any controller
  satisfying the \emph{tightened} constraint
  \begin{equation}
    \Lfho+\Lgho\cdot u
    \;\geq\;
    b(t,\mathbf{y}) + L_{h_1}\delta
    \label{eq:KBLF_tightened}
  \end{equation}
  keeps the trajectory in $\mathcal{S}_{\eps}$ with the same clearance
  floor~\eqref{eq:floor} as in the nominal case for as long as it
  remains in $\mathcal{D}$.
\end{proposition}

\begin{proof}
  Since $\Lgh\equiv0$ regardless of $d$, $\dot h=h_{1}-\alpha(h)$ is
  unaffected. The disturbance enters only through
  $\dot h_{1}=\Lfho+\Lgho\cdot(u+d)$, reducing $\dot\sigma$ by at most
  $L_{h_1}\delta$; by~\eqref{eq:KBLF_tightened},
  $\dot\sigma\geq\bigl(-\mu(\sigma)+L_{h_1}\delta\bigr)
  -L_{h_1}\delta=-\mu(\sigma)$. The proof of
  Theorem~\ref{thm:blf_main} then applies verbatim.
\end{proof}

\begin{corollary}[Explicit disturbance budget without tightening]
\label{cor:robust_explicit}
Let the hypotheses of Theorem~\ref{thm:blf_main} and
Assumption~\ref{ass:disturbance_channel} hold, let $V$ be
\eqref{eq:blf_log} or \eqref{eq:blf_pow} (so that $w(\eps^{+})=0$),
$\bar w\coloneqq\max_{h\in[\eps,h_{c}]}w(h)>0$, and
\begin{equation}
  \delta^{*}\coloneqq\frac{-\mu(-\bar w)}{L_{h_1}} .
  \label{eq:delta_star}
\end{equation}
For $\delta\le\delta^{*}$ define
$\bar s\coloneqq-\mu^{-1}(-L_{h_1}\delta)\in[0,\bar w]$ and let
$h_{\delta}\in(\eps,h_{c}]$ be the largest solution of $w(h)=\bar s$.
If $h(\mathbf{y}(0))\ge h_{\delta}$ and \eqref{eq:margin_compat}
holds, the \emph{untightened} filter $u\in\KBLF$ guarantees
\[
  h\bigl(\mathbf{y}(t)\bigr)\;\geq\;h_{\delta}\;>\;\eps,
  \qquad\forall\,t\in[0,T_{\mathcal{D}}),
\]
where $T_{\mathcal{D}}$ is the first exit time from $\mathcal{D}$.
The floor $h_{\delta}$ equals $h_{c}$ at $\delta=0$ and is
non-increasing in $\delta$.
\end{corollary}

\begin{proof}
With $u\in\KBLF$,
$\dot\sigma\ge F(\sigma)\coloneqq-\mu(\sigma)-L_{h_1}\delta$. $F$ is
continuous and strictly decreasing with $F(-\bar s)=0$, hence $F>0$
on $\{\sigma<-\bar s\}$; repeating Step~3 of
Theorem~\ref{thm:blf_main} with $0$ replaced by $-\bar s$ gives
$\sigma\ge-\bar s$, i.e.\ $\dot h\ge w(h)-\bar s$. Since $w$ is
continuous on $[\eps,h_{c}]$ with
$w(\eps^{+})=w(h_{c})=0\le\bar s\le\bar w$, $h_{\delta}$ exists and
$h_{\delta}>\eps$. As $w\in C^{1}(\eps,\infty)$ and $h_{\delta}$ is
an equilibrium of $\dot z=w(z)-\bar s$, the solution with
$z(0)=h(\mathbf{y}(0))\ge h_{\delta}$ satisfies $z(t)\ge h_{\delta}$
by uniqueness, and the comparison lemma yields
$h(t)\ge z(t)\ge h_{\delta}$.
\end{proof}

Corollary~\ref{cor:robust_explicit} gives an explicit budget
$\delta^{*}$, computable offline from $V$, $\beta$, $h_{c}$, $\mu$,
and $L_{h_1}$.

\begin{revblock}
\textbf{Limits of the robustness results.} 
(i)~\emph{Matched channel only:} disturbances must enter through
$g(t,y_1)$ (Assumption~\ref{ass:disturbance_channel}); disturbances
acting directly on $\dot h$ are not covered.
(ii)~\emph{Known bound:} Proposition~\ref{prop:robust} requires
knowledge of $\delta$ and of $L_{h_1}$, the latter being a worst-case
supremum over $\mathcal{S}_\eps$ that can be conservative, so
$\delta^{*}$ is a sufficient, not a tight, budget.
(iii)~\emph{Beyond $\delta^{*}$} the floor $h_\delta>\eps$ is no longer
guaranteed.

\textit{Unmodeled drift dynamics.} Suppose the drift is $f+\Delta f$
with $\|\Delta f\|\leq\nu$ (unmodeled attention/feedforward effects).
Then $\Lgh\equiv0$ still holds, but now
$\dot h=h_1-\alpha(h)+e$ with $|e|\leq L_h\nu$,
$L_h\coloneqq\sup_{\mathcal{S}_\eps}\|dh\|$, so $u$ has no authority
over $e$, and $\dot h_1$ changes by at most $L'_{h_1}\nu$, $L'_{h_1}\coloneqq\sup_{\mathcal{S}_\eps}\|dh_1\|$. Hence
$\dot\sigma\geq-\mu(\sigma)-\delta_{\mathrm{eff}}$ with
$\delta_{\mathrm{eff}}=(L'_{h_1}+L_\rho L_h)\nu$, where
$L_\rho\coloneqq\sup|\rho'|$ over the region considered, and
$\dot h\geq w(h)-s_{\mathrm{eff}}-L_h\nu$,
$s_{\mathrm{eff}}\coloneqq-\mu^{-1}(-\delta_{\mathrm{eff}})$. A floor
above $\eps$ then exists as in Corollary~\ref{cor:robust_explicit}
provided $s_{\mathrm{eff}}+L_h\nu<\bar w$.
This is only a sketch: $\rho'$ is unbounded as $h\to\eps^{+}$ for the
logarithmic BLF, so the estimate is valid on
$\{h\geq\eps+\kappa\}$ for some $\kappa>0$ and a self-consistent
argument is needed to close it. No theorem is claimed.

\end{revblock}

\subsection{Main Safety Guarantee}
\label{sec:main_result}

\begin{corollary}[BLF safety under the generic surjectivity condition]
  \label{cor:main}
 Let $\eps\in(0,\gamma)$, $h_{c}\in(\eps,\gamma)$, and $h$, $h_{1}$
  as in~\eqref{eq:h}, \eqref{eq:h1_transformer}, with
  Assumptions~\ref{ass:h0} and~\ref{ass:nondeg} holding on an open
  $\mathcal{D}\subseteq\Enk$. Let
  $\mathcal{D}_{\eps}:=\mathcal{D}\cap\{\mathbf{y}:h(\mathbf{y})>\eps\}$.
  Then, for every choice of $(P(t),U(t))\in\bigl(\R^{(n+1)\times(n+1)}\bigr)^{2H}$,
  $t\in\Rgeq$, for which Assumption~\ref{ass:surj} holds on
  $\mathcal{D}$, $\KBLF(t,\mathbf{y})$ is non-empty on
  $\Rgeq\times\mathcal{D}_{\eps}$, and any controller
  $k(t,\mathbf{y})\in\KBLF(t,\mathbf{y})$ applied to an initial
  condition $\mathbf{y}(0)\in\mathcal{D}_{\eps}$ satisfying
  $h_{1}(0,\mathbf{y}(0))\geq\rho(h(\mathbf{y}(0)))$ keeps the
  trajectory in $\mathcal{S}_{\eps}$ for as long as it evolves in
  $\mathcal{D}$. By Proposition~\ref{prop:surj}, this condition is
  generic pointwise in $(t,\mathbf{y})$, but for a particular trained
  model it must still be verified on the certified operating domain
  $\mathcal{D}$. In particular, over that interval,
 \begin{equation}
    \htox\!\bigl(y_{k}(t)\bigr)
    \;\leq\;
    \gamma-\min\bigl(h(\mathbf{y}(0)),h_{c}\bigr)
    \;\leq\;
    \gamma-\eps.
    \label{eq:tox_bound}
  \end{equation}
\end{corollary}

\begin{proof}
  Under Assumption~\ref{ass:surj}, Theorem~\ref{thm:blf_main} gives
  non-emptiness and forward invariance on $\mathcal{D}$, and
  \eqref{eq:tox_bound} follows from~\eqref{eq:floor}.
\end{proof}

Corollary~\ref{cor:main}
strengthens~\cite[Corollary~3.1]{silvestre2025stopping} in two
respects: the toxicity bound is strict
($\htox\leq\gamma-\eps$ rather than $\htox\leq\gamma$),
and~\eqref{eq:blf_decay_rel} certifies exponential restoration of the
clearance level $h_{c}$ whenever it is encroached, beyond merely never
crossing the boundary.

%% ================================================================
\section{Simulation Results}
\label{sec:simulation}
%% ================================================================

We \rev{illustrate} the analysis of Section~\ref{sec:analysis} on a small
two-token ($k=2$), single-head transformer evolving on
$\mathcal{E}^{1}_{W}=S^{1}$ ($W=I_{2}$), with tokens parameterised by
$y=(\cos\theta,\sin\theta)$ and toxicity classifier
$h_{0}(y_{2})=1-y_{2}^{\top}y^{*}$ for a fixed safe direction
$y^{*}$. We set $\gamma=0.5$, $\eps=0.1$, $h_{c}=0.3$, $\beta=2$,
$\mu(s)=s$, and fix key/query/value matrices certified surjective via
$\det\Xi(\mathbf{y})(P,U)\neq0$ (Assumption~\ref{ass:surj}). The
CBF-QP filter of~\cite{silvestre2025stopping}
($\alpha_{1}(h_{1})=h_{1}$) is compared against the BLF-QP filter of
Algorithm~\ref{alg:blf} with the logarithmic BLF~\eqref{eq:blf_log},
both driven by the same nominal controller
$u^{\mathrm{nom}}\equiv0$. 

\begin{figure}[t]
  \centering
\includegraphics[width=0.82\columnwidth, height=0.60\columnwidth, keepaspectratio]{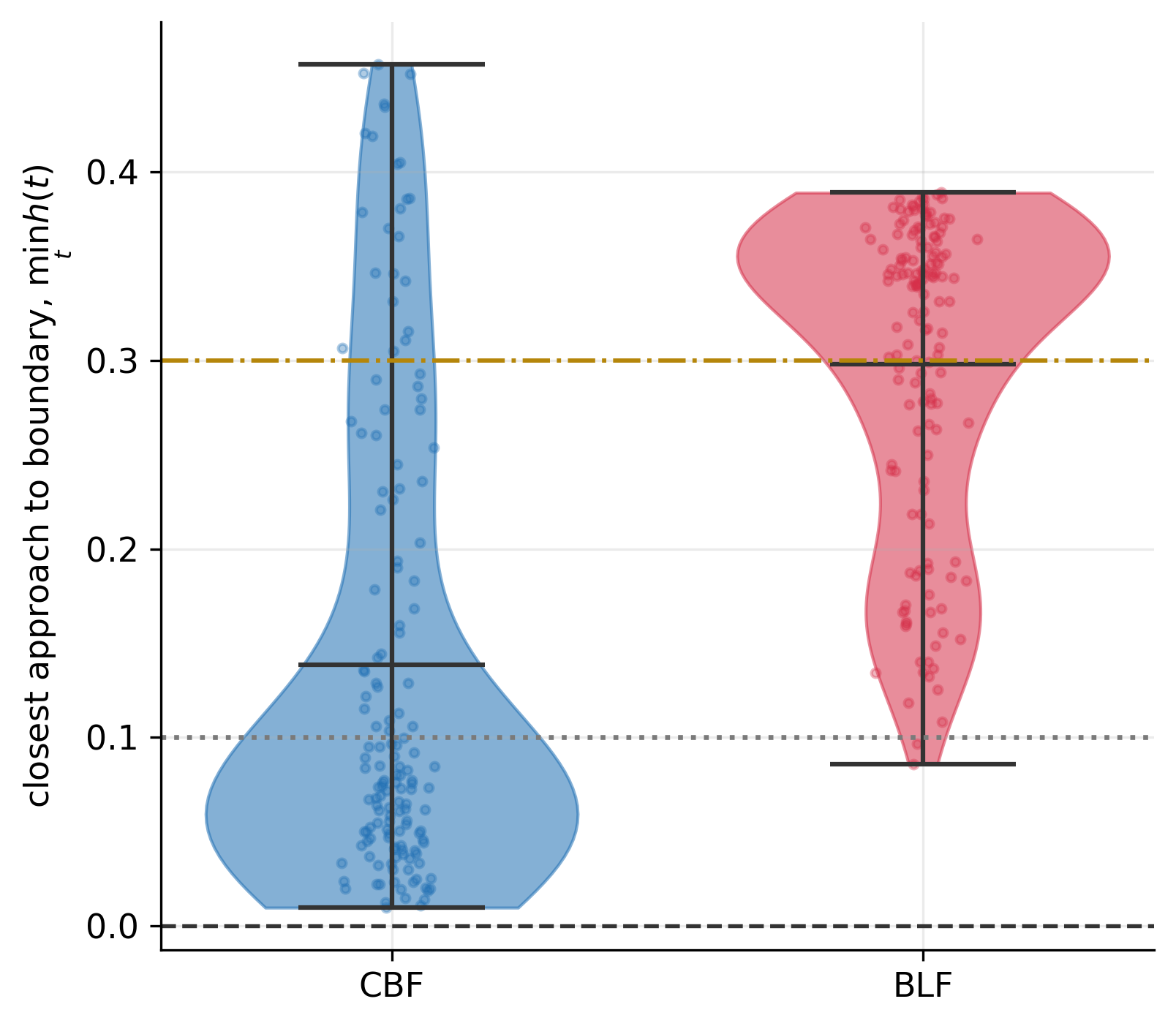}
  \caption{Nominal clearance comparison, $N=150$ trials.
  Distribution of $\min_{t}h(t)$ for the CBF and BLF filters
  ($h=0$ dashed, $h=\eps$ dotted, $h_{c}$ dash-dotted).}
  \label{fig:violin_clearance}
\end{figure}

\begin{figure}[t]
  \centering
 \includegraphics[width=0.85\columnwidth, height=0.60\columnwidth, keepaspectratio]{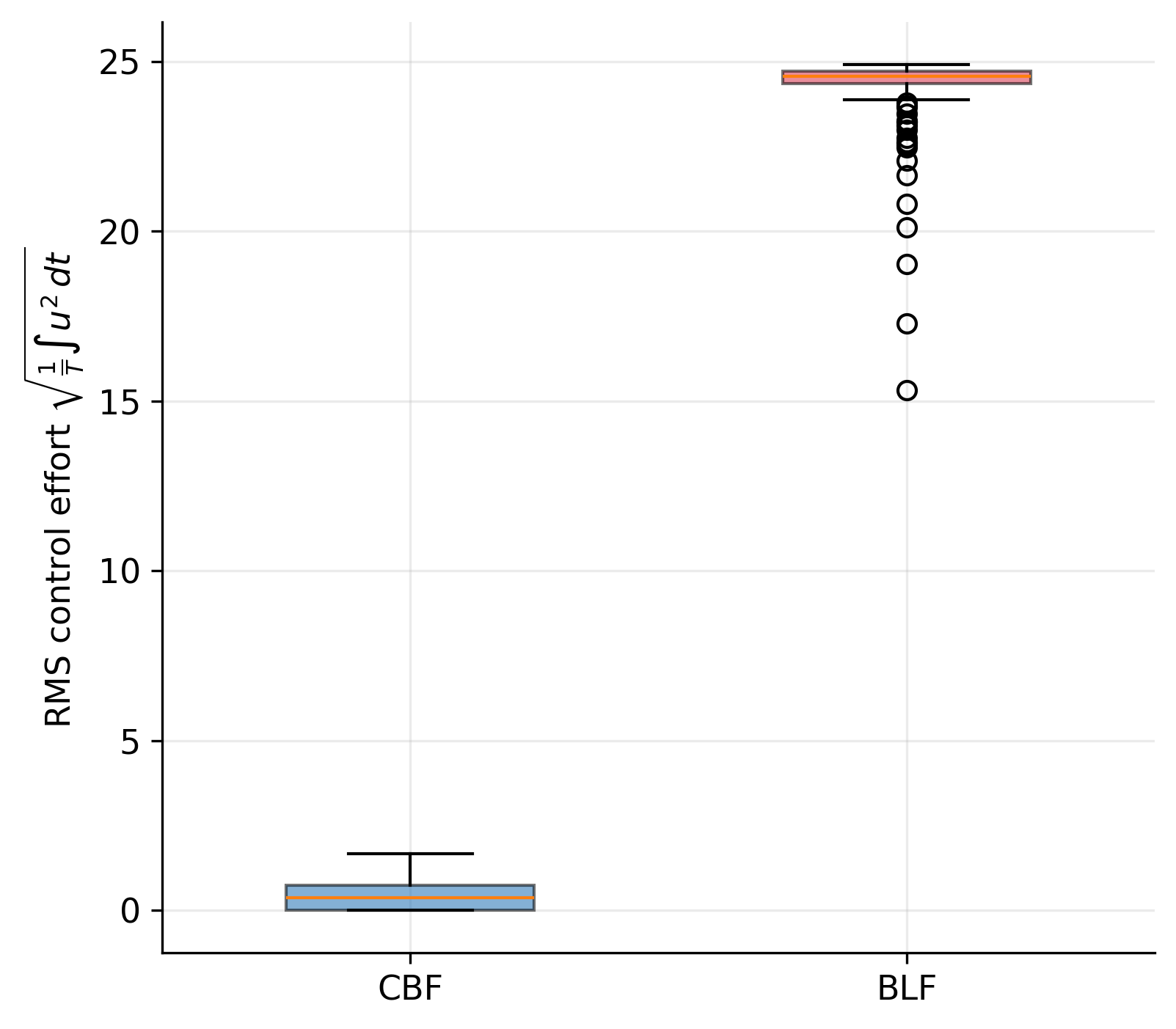}
  \caption{Nominal RMS control effort,
  $(\tfrac{1}{T}\int u^{2}\,dt)^{1/2}$, for the CBF and BLF filters
  over the same $N=150$ trials as Figure~\ref{fig:violin_clearance}.}
  \label{fig:control_effort}
\end{figure}

\textbf{Nominal clearance and control effort.}
Figure~\ref{fig:violin_clearance} reports, over $N=150$ independently
initialised margin-compatible trajectories, the closest approach
$\min_{t}h(t)$ to the toxicity boundary, and
Figure~\ref{fig:control_effort} shows the resulting RMS control effort
$(\tfrac{1}{T}\int u^{2}\,dt)^{1/2}$. The CBF filter repeatedly approaches the safety boundary, with median
closest approach $0.080$, whereas the BLF filter maintains
substantially larger safety margins, with median closest approach
$0.340$; \rev{no trajectory crosses $h=0$, but $1.3\%$ of the undisturbed trials do breach the margin $h=\eps$ (Table~\ref{tab:summary}), so the floor of~\eqref{eq:floor} is \emph{not} met by every trial in this experiment. Theorem~\ref{thm:blf_main} is a continuous-time statement; for the logarithmic BLF the bound $b$ contains $\rho'(h)$, which diverges (logarithmically) as $h\to\eps^{+}$, so a sampled implementation is stiff exactly near the margin and the discrete-time effects of Section~\ref{sec:blf_filter} are a plausible source of these breaches. Confirming this by refining $\Delta t$ is left to future work.} Because $w(h)\leq0$ for $h\geq h_{c}$, the BLF requirement is least demanding at high clearance, and its additional control effort over
the CBF filter is expected to be concentrated in the low-clearance regime
$h<h_{c}$, where the required rate $w(h)$ is largest; for the
logarithmic BLF, $w$ vanishes at both $h=\eps$ and $h=h_{c}$ and
peaks in the interior of the band.

This behavior is exactly what one would expect from a strict-margin
safety filter. The goal is not to minimize control effort regardless of
risk, but to maintain sufficient distance from the toxicity boundary so
that small classification errors, finite integration steps, and control
mismatch do not immediately translate into unsafe text. Indeed, the
filter is intentionally more aggressive near the boundary, and the
resulting increase in RMS effort is the cost of this additional safety
margin. In other words, the controller spends energy precisely where the
decoder is most likely to become unsafe, which is the principled role
of a barrier-based design in safety-critical generation.

\begin{figure}[t]
  \centering
  \includegraphics[width=\columnwidth]{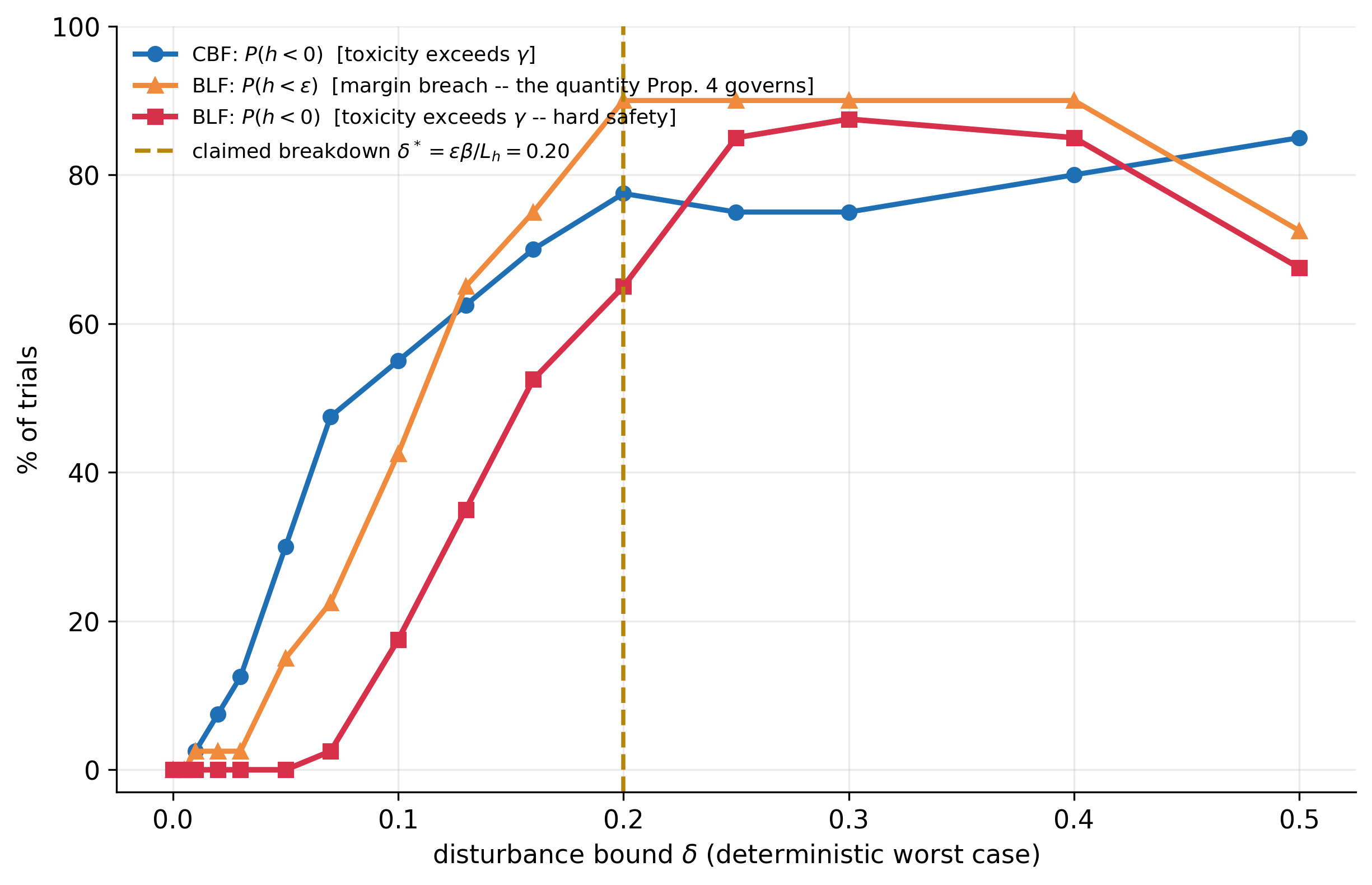}
  \caption{Failure probability vs.\ control-channel disturbance bound
  $\delta$, $N=40$ trials/point: CBF $P(h<0)$, BLF $P(h<\eps)$, BLF
  $P(h<0)$. Vertical line: disturbance budget
   $\delta^{\ast}=-\mu(-\bar w)/L_{h_1}$
  (Corollary~\ref{cor:robust_explicit}).}
  \label{fig:disturbance_sweep}
\end{figure}

\textbf{Disturbance sweep.}
Figure~\ref{fig:disturbance_sweep} sweeps the control-channel
disturbance bound $\delta$ in~\eqref{eq:perturbed} under an adversarial,
sup-norm-bounded perturbation, reporting over $N=40$ trials per value of
$\delta$ the empirical probability of the CBF bare-safety failure $h<0$,
the BLF margin failure $h<\eps$, and the BLF bare-safety failure $h<0$.
The \emph{untightened} BLF filter is used, so the data are best read as a
robustness diagnostic rather than a universal operational certificate: the
sweep clarifies the practical effect of disturbance on the margin, but the
continuous-time proof still operates in the idealized model and with a
smooth classifier.

\begin{revblock}
\begin{table}[t]
\centering
\caption{Summary of the nominal and disturbed-case experiments. The BLF filter trades extra control effort for a larger buffer from the toxicity boundary, while the CBF filter remains closer to the threshold. The bold row highlights the strict-margin metric that the theory targets.}
\label{tab:summary}
\resizebox{\columnwidth}{!}{%
\begin{tabular}{lcc}
\toprule
Metric & CBF (prior work) & BLF (this work)\\
\midrule
$P(h<0)$, no disturbance & $0.0\%$ & $0.0\%$\\
$P(h<\eps)$, no disturbance & n/a & $1.3\%$\\
Mean closest approach to boundary & $0.138$ & $0.298$ ($+116\%$)\\
Median closest approach to boundary & $0.080$ & $0.340$\\
Mann--Whitney $p$-value (clearance shift) & --- & $2.7\times10^{-24}$\\
$P(h<0)$ at $\delta=0.10$ (adversarial) & $55\%$ & $18\%$\\
$\bm{P(h<\eps)}$ \textbf{at} $\bm{\delta=0.10}$ \textbf{(strict-margin metric)} & n/a & $\bm{42\%}$\\
$P(h<0)$ at $\delta=0.20$ (adversarial) & $78\%$ & $65\%$\\
RMS control effort & $0.43$ & $24.2$\\
\bottomrule
\end{tabular}}
\end{table}
\end{revblock}

\begin{revblock}
\textbf{Scope and limitations of the experiments.} The experiments use
a single, very small model ($k=2$ tokens on $S^{1}$, one head, one fixed
certified parameter set), a one-parameter-family linear-in-token
classifier $h_0$, one setting of $(\gamma,\eps,h_c,\beta,\mu)$, and a
zero nominal controller. They \emph{illustrate} the theory in a setting where all
assumptions can be verified; they do not establish the behaviour of
realistic LLMs, learned toxicity classifiers, or non-trivial nominal
controllers, and no claim about practical LLM safety is made. The
experiments validate the strict-margin mechanism and its disturbance
trends within the certified model class; they are not intended as
evidence of safety performance on deployed pretrained LLMs. Larger
studies (more tokens, random certified $(P,U)$, parameter sweeps,
non-zero $u^{\mathrm{nom}}$, and a pretrained model with a smooth
surrogate classifier) are future work.
\end{revblock}

\section{Conclusion}
\label{sec:conclusion}

This paper introduced a Barrier Lyapunov Function safety filter for
transformer text generation that keeps the toxicity score a guaranteed
distance below the threshold, rather than merely below it. For the
continuous-time transformer model, the filter is a small quadratic
program that stays feasible whenever the deployed attention parameters
satisfy the generic surjectivity and classifier-nondegeneracy
conditions on the certified operating domain, never violates the
margin from margin-compatible initial conditions, and restores a
chosen clearance level exponentially
whenever it is encroached. Its admissible trajectories satisfy the exponential second-order CBF
condition as the margin vanishes, and it tolerates control-channel
disturbances up to an explicit sufficient budget.
These are theoretical guarantees for an idealized model and a smooth
classifier. The small simulation illustrates the clearance gain but does
not yet confirm the disturbance budget, and nothing is claimed for
deployed language models. Future work includes smooth surrogates for
black-box classifiers, drift uncertainty, input constraints,
sampled-data guarantees, larger experiments on pretrained models, and
extensions to multimodal models.

	\bibliographystyle{ieeetr}   % or plain, unsrt, alpha, etc.
	\bibliography{reference}

@inproceedings{silvestre2025stopping,
  title={Stopping LLMs from Going Rogue: A Control Barrier Approach to Text Generation},
  author={Silvestre, Jo{\~a}o Pedro and Abella, {\'A}lvaro Rodr{\'\i}guez and Tabuada, Paulo},
  booktitle={2025 IEEE 64th Conference on Decision and Control (CDC)},
  pages={151--156},
  year={2025},
  organization={IEEE}
}

@article{vaswani2017attention,
  title={Attention is all you need},
  author={Vaswani, Ashish and Shazeer, Noam and Parmar, Niki and Uszkoreit, Jakob and Jones, Llion and Gomez, Aidan N and Kaiser, {\L}ukasz and Polosukhin, Illia},
  journal={Advances in neural information processing systems},
  volume={30},
  year={2017}
}

@article{radford2018improving,
  title={Improving language understanding by generative pre-training},
  author={Radford, Alec and Narasimhan, Karthik and Salimans, Tim and Sutskever, Ilya and others},
  year={2018},
  publisher={San Francisco, CA, USA}
}

@article{achiam2023gpt,
  title={Gpt-4 technical report},
  author={Achiam, Josh and Adler, Steven and Agarwal, Sandhini and Ahmad, Lama and Akkaya, Ilge and Aleman, Florencia Leoni and Almeida, Diogo and Altenschmidt, Janko and Altman, Sam and Anadkat, Shyamal and others},
  journal={arXiv preprint arXiv:2303.08774},
  year={2023}
}

@article{guo2025deepseek,
  title={Deepseek-r1: Incentivizing reasoning capability in llms via reinforcement learning},
  author={Guo, Daya and Yang, Dejian and Zhang, Haowei and Song, Junxiao and Wang, Peiyi and Zhu, Qihao and Xu, Runxin and Zhang, Ruoyu and Ma, Shirong and Bi, Xiao and others},
  journal={arXiv preprint arXiv:2501.12948},
  year={2025}
}

@inproceedings{gehman2020realtoxicityprompts,
  title={Realtoxicityprompts: Evaluating neural toxic degeneration in language models},
  author={Gehman, Samuel and Gururangan, Suchin and Sap, Maarten and Choi, Yejin and Smith, Noah A},
  booktitle={Findings of the association for computational linguistics: EMNLP 2020},
  pages={3356--3369},
  year={2020}
}

@inproceedings{oh2023token,
  title={Token-wise decomposition of autoregressive language model hidden states for analyzing model predictions},
  author={Oh, Byung-Doh and Schuler, William},
  booktitle={Proceedings of the 61st Annual Meeting of the Association for Computational Linguistics (Volume 1: Long Papers)},
  pages={10105--10117},
  year={2023}
}

@inproceedings{madsen2024self,
  title={Are self-explanations from Large Language Models faithful?},
  author={Madsen, Andreas and Chandar, Sarath and Reddy, Siva},
  booktitle={Findings of the Association for Computational Linguistics: ACL 2024},
  pages={295--337},
  year={2024}
}

@article{suau2024whispering,
  title={Whispering experts: Neural interventions for toxicity mitigation in language models},
  author={Suau, Xavier and Delobelle, Pieter and Metcalf, Katherine and Joulin, Armand and Apostoloff, Nicholas and Zappella, Luca and Rodr{\'\i}guez, Pau},
  journal={arXiv preprint arXiv:2407.12824},
  year={2024}
}

@inproceedings{luong2024realistic,
  title={Realistic evaluation of toxicity in large language models},
  author={Luong, Tinh and Le, Thanh-Thien and Ngo, Linh and Nguyen, Thien},
  booktitle={Findings of the Association for Computational Linguistics: ACL 2024},
  pages={1038--1047},
  year={2024}
}

@inproceedings{chao2025jailbreaking,
  title={Jailbreaking black box large language models in twenty queries},
  author={Chao, Patrick and Robey, Alexander and Dobriban, Edgar and Hassani, Hamed and Pappas, George J and Wong, Eric},
  booktitle={2025 IEEE Conference on Secure and Trustworthy Machine Learning (SaTML)},
  pages={23--42},
  year={2025},
  organization={IEEE}
}

@misc{geshkovski2025mathematicalperspectivetransformers,
      title={A mathematical perspective on Transformers}, 
      author={Borjan Geshkovski and Cyril Letrouit and Yury Polyanskiy and Philippe Rigollet},
      year={2025},
      eprint={2312.10794},
      archivePrefix={arXiv},
      primaryClass={cs.LG},
      url={https://arxiv.org/abs/2312.10794}, 
}

@inproceedings{
karagodin2024clustering,
title={Clustering in Causal Attention Masking},
author={Nikita Karagodin and Yury Polyanskiy and Philippe Rigollet},
booktitle={The Thirty-eighth Annual Conference on Neural Information Processing Systems},
year={2024},
url={https://openreview.net/forum?id=OiVxYf9trg}
}

@inproceedings{
abella2025consensus,
title={Consensus Is All You Get: The Role of Attention in Transformers},
author={{\'A}lvaro Rodr{\'\i}guez Abella and Jo{\~a}o Pedro Silvestre and Paulo Tabuada},
booktitle={Forty-second International Conference on Machine Learning},
year={2025},
url={https://openreview.net/forum?id=XROl6J4wjX}
}

@misc{
cheng2025linearly,
title={Linearly Controlled Language Generation with Performative Guarantees},
author={Emily Cheng and Marco Baroni and Carmen Amo Alonso},
year={2025},
url={https://openreview.net/forum?id=t8ctvylFn7}
}

@misc{
miyaoka2025cbfllm,
title={{CBF}-{LLM}: Safe Control for {LLM} Alignment},
author={Yuya Miyaoka and Masaki Inoue},
year={2025},
url={https://openreview.net/forum?id=fvo6q86NKG}
}

@ARTICLE{7782377,
  author={Ames, Aaron D. and Xu, Xiangru and Grizzle, Jessy W. and Tabuada, Paulo},
  journal={IEEE Transactions on Automatic Control}, 
  title={Control Barrier Function Based Quadratic Programs for Safety Critical Systems}, 
  year={2017},
  volume={62},
  number={8},
  pages={3861-3876},
  doi={10.1109/TAC.2016.2638961}}

@INPROCEEDINGS{7524935,
  author={Nguyen, Quan and Sreenath, Koushil},
  booktitle={2016 American Control Conference (ACC)}, 
  title={Exponential Control Barrier Functions for enforcing high relative-degree safety-critical constraints}, 
  year={2016},
  volume={},
  number={},
  pages={322-328},
  doi={10.1109/ACC.2016.7524935}}

@ARTICLE{9516971,
  author={Xiao, Wei and Belta, Calin},
  journal={IEEE Transactions on Automatic Control}, 
  title={High-Order Control Barrier Functions}, 
  year={2022},
  volume={67},
  number={7},
  pages={3655-3662},
  doi={10.1109/TAC.2021.3105491}}

@INPROCEEDINGS{8581068,
  author={Sachan, Kapil and Padhi, Radhakant},
  booktitle={2018 15th International Conference on Control, Automation, Robotics and Vision (ICARCV)}, 
  title={Barrier Lyapunov Function Based Output-constrained Control of Nonlinear Euler-Lagrange Systems}, 
  year={2018},
  volume={},
  number={},
  pages={686-691},
  doi={10.1109/ICARCV.2018.8581068}}

@INPROCEEDINGS{1583507,
  author={Ngo, K.B. and Mahony, R. and Zhong-Ping Jiang},
  booktitle={Proceedings of the 44th IEEE Conference on Decision and Control}, 
  title={Integrator Backstepping using Barrier Functions for Systems with Multiple State Constraints}, 
  year={2005},
  volume={},
  number={},
  pages={8306-8312},
  doi={10.1109/CDC.2005.1583507}}

@misc{mityagin2015zerosetrealanalytic,
      title={The Zero Set of a Real Analytic Function}, 
      author={Boris Mityagin},
      year={2015},
      eprint={1512.07276},
      archivePrefix={arXiv},
      primaryClass={math.CA},
      url={https://arxiv.org/abs/1512.07276}, 
}
\end{document}